%% file: main_6pages.tex
\documentclass[conference]{IEEEtran}
\IEEEoverridecommandlockouts

\usepackage{cite}
\usepackage{amsmath,amssymb,amsfonts}
\usepackage{algorithmic}
\usepackage{graphicx}
\usepackage[toc, acronym, nopostdot, nonumberlist]{glossaries}
\input{glossary}
\usepackage{textcomp}
\usepackage{balance}
\usepackage{multicol}
\usepackage{multirow}
\usepackage{xcolor}
\usepackage{fontawesome}
\usepackage[utf8]{inputenc}

\usepackage[most]{tcolorbox}

\definecolor{promptblue}{RGB}{225,240,255}
\definecolor{responsegreen}{RGB}{230,245,230}

\usepackage[table]{xcolor}
\usepackage{tabularx}

\definecolor{promptblue}{RGB}{225,240,255}
\definecolor{responsegreen}{RGB}{230,245,230}

\def\BibTeX{{\rm B\kern-.05em{\sc i\kern-.025em b}\kern-.08em
    T\kern-.1667em\lower.7ex\hbox{E}\kern-.125emX}}
\usepackage{lipsum}

\title{Agentic Workflows for Resolving Conflict Over Shared Resources: A Power Grid Application
\thanks{\IEEEauthorrefmark{1}These authors contributed equally to this work.}
}  

\author{\IEEEauthorblockN{Shiva Poudel\IEEEauthorrefmark{1}, Thiagarajan Ramachandran\IEEEauthorrefmark{1}, Orestis Vasios, and Andrew P. Reiman \\}
\IEEEauthorblockA{Pacific Northwest National Laboratory, Richland, WA 99354 \\
\faEnvelopeO\ \{shiva.poudel, thiagarajan.ramachandran, orestis.vasios, andrew.reiman\}@pnnl.gov 
\vspace{-0.25 cm}}
}

\renewcommand{\footnoterule}{%
\kern -3pt
\hrule width 0.49 \textwidth height 0.5pt
\kern 1pt
}

\begin{document}
\begingroup
\allowdisplaybreaks

\maketitle

\begin{abstract}
The increasing use of LLM-based agents to support decision-making and control across diverse domains motivates the need for systematic deconfliction of their proposed actions. We present a deconfliction framework for coordinating multiple agents that formally encapsulate individual applications, each proposing potentially conflicting actions over shared resources. 
Conflicts are resolved through three deconfliction modes: bilateral negotiation, structured mediation, and procedural (deterministic) deconfliction. We define design principles for large language model-based client agents, including a chain-of-thought style reasoning process, and introduce an iterative weighted-consensus mechanism that does not require the applications themselves to solve optimization problems. The framework is domain agnostic and supports both numeric and non-numeric decisions. Its performance is demonstrated on a power distribution use case with conflicting \gls{adms} applications for cost optimization and resilience, coordinating diesel generators and battery energy storage systems.
\end{abstract}



\section{Introduction}


Modern infrastructures and software ecosystems increasingly rely on collections of semi-autonomous agents that sense, decide, and act on shared environments. These agents are now embedded in domains such as electric power systems, air traffic management, logistics, finance, and multi-robot systems, where they continuously propose actions over shared resources and constraints \cite{dong2023survey}. As recent advances in \gls{llm} have enabled more capable autonomous behavior, \gls{llm}-based agents are being introduced as higher-level decision-makers that coordinate, supervise, or augment these semi-autonomous components \cite{zhang2024application, mathes2025collaborative, aghaee2025rb, li2025review}. 
As their number and sophistication grow, they frequently conflict in their proposed actions, resource usage, or objectives, which can degrade performance, violate safety or regulatory requirements, and erode trust in autonomous decision-making.
This motivates principled deconfliction mechanisms that can detect, reason about, and resolve incompatible agent behaviors while preserving as much local autonomy as possible \cite{reiman2023app}.

Across domains, conflicts among agents typically arise from a small number of recurring causes: multi-objective or misaligned goals, competition for shared resources, heterogeneous or partial information, fragmented governance or authority, and timing or synchronization issues \cite{sapkota2025ai}. 
Traditional conflict resolution approaches often assume perfect information, centralized control, or highly structured agent architectures, which simplifies analysis but limits applicability in large-scale, heterogeneous systems \cite{pallottino2004decentralized}. For example, many classical schemes presuppose a single coordinator with access to the global system state and all agents’ objectives, enabling it to compute and impose conflict-free joint actions \cite{rizk2018decision}. Similarly, foundational models of multi-agent decision making and negotiation frequently rely on common knowledge of system dynamics and other agents’ preferences or policies, assumptions that break down when agents are independently designed, partially informed, or subject to privacy constraints~\cite{olfati2004consensus}.

Given the widespread deployment of \gls{llm}-based autonomous agents and their increasing tendency to conflict over shared resources and objectives, there is a critical need for deconfliction mechanisms tailored to these systems' unique characteristics of natural language reasoning, distributed decision-making under uncertainty, and privacy-preserving operation, where agents may not fully disclose their objectives. To address this gap, we propose a framework for deconfliction based on \cite {reiman2023app} specifically designed to enable \gls{llm}-based agents resolve conflicts over shared resources while preserving their autonomous capabilities and confidential information. The specific contributions of the papers are:
\begin{enumerate}
    \item Introduce a domain-agnostic formulation of application deconfliction that treats applications as abstract agents with proposals over shared resources, and instantiates three concrete deconfliction modes\textemdash bilateral negotiation, structured mediator, and procedural (deterministic) deconfliction\textemdash each with clearly specified workflows and design principles.
    \item Present general design principles for deconfliction agents and provide a formal formulation of the proposed \gls{cot}, and characterize the agents’ reasoning structure by outlining their internal decision flow.
    \item Compare bilateral negotiation, structured mediation, and procedural deconfliction as mechanisms for resolving conflicts among agents using a power grid case study, evaluating each across consensus trajectory, resolution vector, and deviation from both agent-optimal and centroid-based baselines.
\end{enumerate}

\section{Agents for Client Applications}
Modern cyber-physical systems increasingly involve multiple applications operating concurrently and producing control actions that may conflict over shared resources. For example, a \gls{der} management application may dispatch a battery to maximize self-consumption while a grid operator application simultaneously attempts to use the same asset for frequency regulation. Wrapping applications like these with LLM agents, hereby referred to as \textit{client agents}, enables them to reason about and negotiate such conflicts without requiring conflict resolution logic to be implemented for every possible interaction. This design offers several practical advantages:
\begin{itemize}
\item \textbf{Integration Complexity:} Client agents provide a common natural language interface for deconfliction across diverse applications that may otherwise have incompatible communication protocols or conflicting control hierarchies, reducing the overhead of integrating new applications into the deconfliction framework. Agents do not require access to the internal logic of applications that may be proprietary. The agent interacts with an application through its native inputs and outputs and is able to treat the underlying application as another tool.
\item \textbf{Configuration Complexity:} The policies governing deconfliction behavior can be defined and updated through straightforward prompts or instruction modifications, lowering configuration complexity and enabling reconfiguration under changing operational requirements. 
\item \textbf{Enables Application Diversity:} Due to their natural language processing capabilities, client agents can interface with a broad range of decision-making processes and support both quantitative and qualitative decision making.
\end{itemize}
In order to enable client agents to negotiate in a structured and principled manner, we establish a set of general design principles that characterize well-behaved client agents: 

\begin{itemize}
\item \textbf{Separation of Concerns:} The client agent should be decoupled from the underlying application logic. The client agent is responsible for negotiation and conflict resolution and should use the application as a tool for generating and testing candidate actions given a set of inputs. This separation ensures that the agents can be applied to arbitrary applications with minimal modifications.

\item \textbf{Preference Communication under Privacy:} Agents must be able to participate in deconfliction
without revealing their underlying objective functions or private operational constraints. This is particularly important in multi-stakeholder environments where applications may be owned or operated by different entities.

\item \textbf{Pursuit of Objective:} Agents should pursue their best individual outcome in a manner consistent with a multi-agent context. Agents need to balance their internal objectives with the incentive for compromise imposed by the deconfliction framework.
\end{itemize}
These principles can then be used to inform the construction of the \gls{cot} prompts that govern agent behavior.

\section{Deconfliction Formulation}
The application deconfliction framework described in \cite{reiman2023app} is summarized and extended for LLM-based agents in this section. The application deconfliction framework was originally conceived as a way to improve integration and performance of applications developed by different third parties hosted on distribution operations platforms (e.g., advanced distribution management systems). In the application deconfliction framework, device setpoints produced by all applications are routed through a deconfliction pipeline that determines final values for all setpoints prior to dispatch. This architecture allows application with overlapping decision space to be designed and deployed. This approach eliminates the need for applications to be designed specifically to work together and opens the door to improved system performance by allowing more applications to partially, conditionally, or fully influence more controllable setpoints.

Several methods can be used to resolve conflicting setpoints produced by different applications. These range from traditional hierarchical or modal access to specific setpoints by specific applications to multi-objective optimization, multi-criteria decision-making, and game-theoretic methods that enable win-win outcomes for all applications. When applications are protected from disclosing internal logic or objectives, game theoretic methods offer a way to incentivize applications to demonstrate cooperative behavior. In \cite{reiman2024gametheoretic}, a game was introduced that iteratively computes a weighted centroid of requested setpoints, where weight factors are determined based on how much flexibility each application demonstrates, rewarding applications that compromise on setpoints that have relatively little effect on their internal objective. 

In this work, we have adapted application deconfliction and the game theoretic approach in \cite{reiman2024gametheoretic} for LLM-based agents. LLM-based agents can implement a wider range of decision-making processes (e.g., qualitative decision-making in addition to the optimization and heuristic methods available to traditional process-based applications and available to LLM-based agents through tools). LLM-based agents can also exchange text-based justification of setpoint requests, enabling different modes of mediation and even direct agent-to-agent negotiation as described in Section IV.

\begin{figure}[h]
    \centering
    \includegraphics[width=0.55\linewidth]{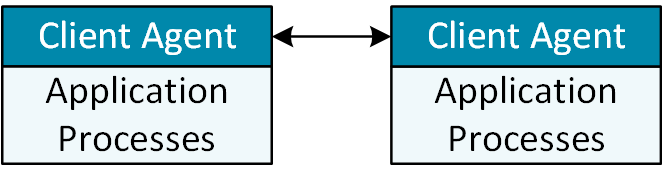}
    \vspace{-0.2 cm}
    \caption{Bilateral negotiation for agent deconfliction.}
    \vspace{-0.25 cm}
    \label{fig:bn_workflow}
\end{figure}

\begin{figure}[t]
    \centering
    \includegraphics[width=0.85\linewidth]{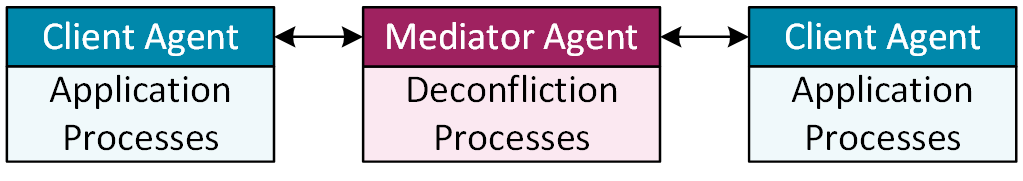}
    \vspace{-0.2 cm}
    \caption{Structured mediator-based deconfliction of agents.}
    \label{fig:ms_workflow}
\end{figure}

\begin{figure}[t]
    \centering
    \includegraphics[width=0.85\linewidth]{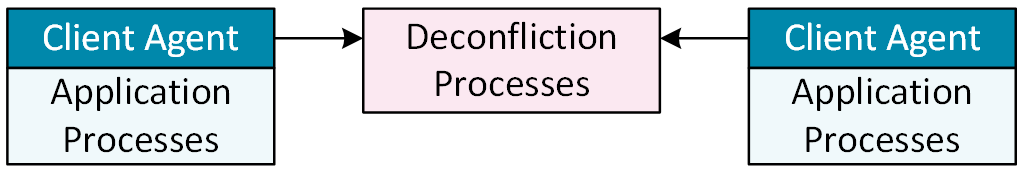}
    \vspace{-0.2 cm}
    \caption{Procedure-based deconfliction of agents.}
    \vspace{-0.5 cm}
    \label{fig:ps_workflow}
\end{figure}

\section{Deconfliction Modes for Agents}

There are different modes of agent deconfliction and each of them is described here.


\subsection{Bilateral Negotiation}
Figure \ref{fig:bn_workflow} illustrates the bilateral negotiation framework. Each application is paired with an agent that represents the application’s objectives, constraints, and preferred control setpoints in a negotiation. The agents interact directly with one another, exchanging offers and adjusted counter-offers while iteratively adjusting their proposals to account for conflicts over shared resources. Through this process, the agents converge toward mutually acceptable agreements that balance competing priorities, while still preserving the autonomy of each application. 

\subsection{Structured Mediator}
Figure \ref{fig:ms_workflow} illustrates the mediated negotiation framework used for agent deconfliction. Each application is represented by its own agent, and negotiation is coordinated through a dedicated mediator agent. The mediator collects proposed setpoints from the agents, evaluates conflicts using a structured deconfliction process, and returns revised compromise proposals. This approach preserves agent autonomy while providing a disciplined mechanism for coordinating competing application demands through a neutral intermediary.

The structured deconfliction process is shown in Fig. \ref{fig:deconfliction_process}. The mediator-guided deconfliction process resolves conflicts among agents through an iterative consensus mechanism. 

\begin{figure}
    \centering
    \includegraphics[width=0.95\linewidth]{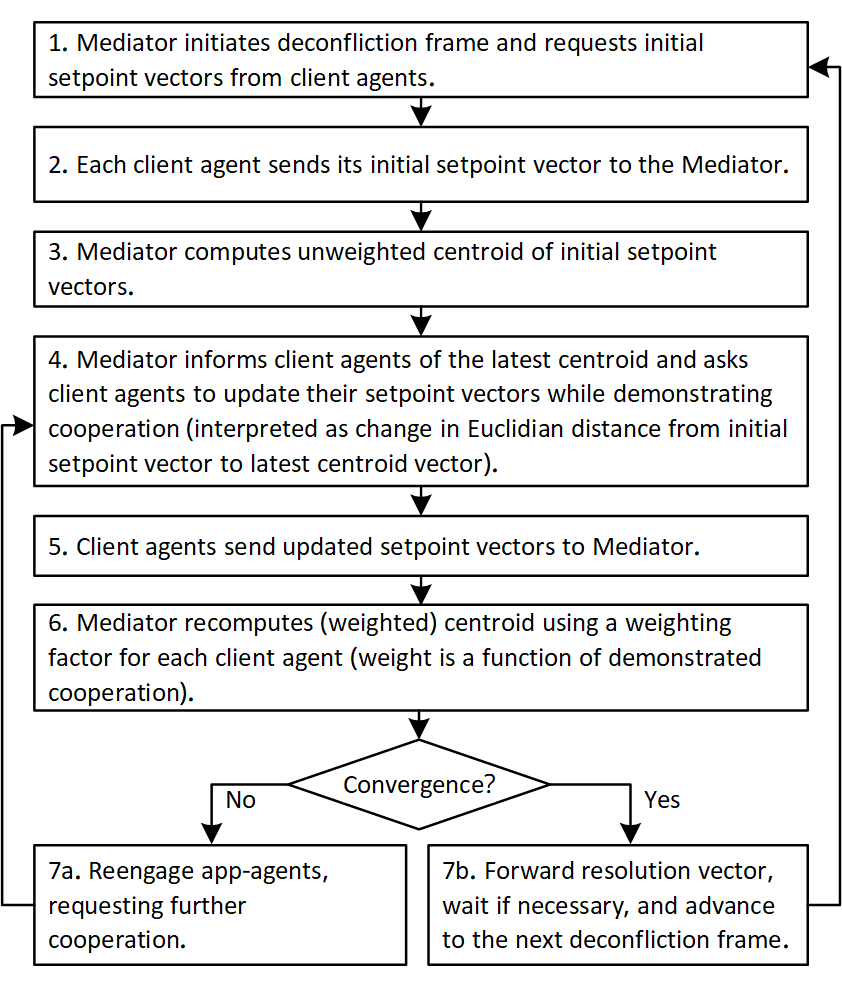}
    \vspace{-0.25 cm}
    \caption{Deconfliction process: weighted consensus method}
    \label{fig:deconfliction_process}
    \vspace{-0.5 cm}
\end{figure}


\subsection{Procedural Deconfliction}
The procedural deconfliction mode follows a workflow similar to the structured mediator, but instead of a mediator agent, it uses a deterministic deconflictor (Fig. \ref{fig:ps_workflow}) that performs weight computation and centroid evaluation as described in \cite{reiman2024gametheoretic} adapted for LLM-based agents. For each iteration $k$, weighted centroids $c_{k}$ are computed:
\begin{align}
    c_{k}=\frac{\sum_{i}{(w_{i,k} \cdot x_{i,k})}}{\sum_{i}{w_{i,k}}}
\end{align}
where $x_{i,k}$ is the setpoint vector proposed by agent $i$ in iteration $k$ and $w_{i,k}$ is the weight factor assigned to agent $i$ in iteration $k$. In the initial iteration, $k=0$, $w_{i,0}=1$; for subsequent iterations $k>0$ weight factors are calculated based on the flexibility demonstrated by the agent:
\begin{align}
    w_{i,k} =  \frac {\langle {x_{i,0},c_{k-1}} \rangle} {\langle {x_{i,k},c_{k-1}} \rangle} 
\end{align}
where $\langle \cdot \rangle$ represents the Euclidean distance between a pair of vectors. If any $x_{i,k}=c_{k-1}$ for any agent, that agent's weighting factor is set to infinity and no additional iterations are performed. This formulation differs from \cite{reiman2024gametheoretic} in the calculation of $w$. This supports agents that use heuristic or qualitative methods to play the game, ultimately to near convergence, but can produce unpredictable results agents demonstrate extreme flexibility. To alleviate this concern, we design our agents to balance flexibility with their internal ideal solution and conclude the game after a predetermined number of iterations.


\section{Deconfliction for Advanced Distribution Operations}
In this section, we describe a power grid example where \gls{llm}-based client agents represent distribution operations applications whose objectives are fundamentally in tension.

Conflict in power distribution operations arises as a direct consequence of the rapid deployment of \gls{adms} and \gls{derms} applications, each issuing control actions to overlapping sets of field devices in pursuit of distinct objectives such as affordability and reliability. 


\subsection{Example Applications}
We consider a distribution system equipped with diesel generators (DGs), photovoltaic units (PVs), and \gls{bess}. These devices are controlled by two distribution operation applications: a cost application and a resilience application. Let $P_{\text{DG}}^i$ denote the active power output of DG $i$, $P_{\text{PV}}^j$ the active power output of PV unit $j$, and $P_{\text{ESS}}^k$ the active power of ESS unit $k$ (with $P_{\text{ESS}}^k < 0$ representing charging). 
\subsubsection{Cost Optimization Application}

The cost application aims to minimize the operational cost of supplying demand, considering time-varying grid prices, diesel generation cost, and \gls{bess} operation. Let $\lambda_{\text{grid}}^t$ denote the grid electricity price at time $t$, $\lambda_{\text{grid}}^{\text{avg}}$ the average grid price over the horizon, and $c_{\text{DG}}$ the DG operational cost coefficient (\$/kWh). The instantaneous cost function is
\begin{align}\label{cost}
    J_{\text{cost}}^{t}
    &= \lambda_{\text{grid}}^{t}\left(
        - \sum_{i} P_{\text{DG}}^{i,t}
        - \sum_{j} P_{\text{PV}}^{j,t}
        - \sum_{k} P_{\text{ESS}}^{k,t}
      \right) \nonumber\\
    &\quad + c_{\text{DG}} \sum_{i} P_{\text{DG}}^{i,t}
      + \lambda_{\text{grid}}^{\text{avg}} \sum_{k} P_{\text{ESS}}^{k,t},
\end{align}
and the application minimizes the total cost $\sum_t J_{\text{cost}}(t)$ over 24 hour subject to network and device constraints.

\subsubsection{Resilience Application}

The resilience application focuses on preserving reserves by coordinating DGs and \gls{bess}. It seeks to minimize DG active power output and promote \gls{bess} charging. The resilience objective is defined as
\begin{equation}\label{resilience}
    J_{\text{resilience}}^{t}
    = \sum_{i} P_{\text{DG}}^{i,t}
      + \sum_{k} P_{\text{ESS}}^{k,t}
\end{equation}
Minimizing $J_{\text{resilience}}^t$ encourages lower DG active power, negative \gls{bess} power (charging or reserve buildup), thereby enhancing overall system resilience.

Each of these two applications and their associated processes is represented by an agent that acts on its behalf. 
These client agents are designed according to the principles described in Section II. 
The agentic workflow uses Claude 4.5 Sonnet as the underlying \gls{llm}, accessed via the OpenAI Agents SDK. Two independent client agents are instantiated using the Agent class, with custom prompts and specialized function tools. 

\begin{table}[t]
\centering
\caption{\gls{der} Locations and Capacities in the Test System.}
\label{table}
\setlength{\tabcolsep}{3pt}
\begin{tabular}{cccc}
\hline
Bus & Type of DER & $p_{\mathrm{max},3\phi}$ ($\mathrm{kW}$) & Capacity ($\mathrm{kWh}$) \\
\hline
47  & DG    & 300   & -     \\
48  & PV    & 300   & -     \\
49  & DG    & 300   & -     \\
65  & PV    & 300   & -     \\
76  & PV    & 300   & -     \\
48  & BESS  & 150   & 1200  \\
76  & BESS  & 187.5 & 1500  \\
\hline
\end{tabular}
\vspace{-0.35 cm}
\label{table:mod_123}
\end{table}

\subsection{Tools Available for Agents}
The following tools are made available for client agents to facilitate the submission of initial proposals as well as counterproposals during negotiations.
\subsubsection{Optimal Setpoints}
Equations \eqref{cost} and \eqref{resilience} enable client agents to evaluate their desired setpoints for a given time.
\subsubsection{Balanced Compromise}
The deconfliction process requires client agents to compromise on their desired setpoints. Because this compromise may adversely affect their private objectives, agents must strike a balance. To support this, agents are given access to a tool that computes an optimal compromise based on the current consensus and agent-defined flexibility. Given the consensus setpoint $x_{con}$, agent $i$ with a private objective $J_i$, the tool that solves the following optimization problem is:
\begin{align}
&\min_{x}J_i(x)\\
&\text{s.t}~~x \in B_{\alpha_i}(x),~x \in \mathcal{F}
\end{align}
where $\mathcal{F}$ is the set of feasible setpoints and $B_{\alpha_i}(x)$ is a closed ball centered around $x$ with radius $\alpha_i$. $\alpha_i$ represents the flexibility parameter chosen by the agent that determines the degree of compromise. The agent selects $\alpha_i \in [0, d_i]$ where 0 represents no compromise and $d_i$
is the current distance between the consensus setpoint and the agent's desired value. The solution to the optimization problem will move the agents closer to the consensus by $d_i - \alpha_i$ while reducing the adverse impact of the compromise on the objective $J_i$. Note that this is not a unique approach to determining the compromise, but it will guarantee convergence. 

\begin{figure}[t]
    \centering
    \includegraphics[width=0.99\linewidth]{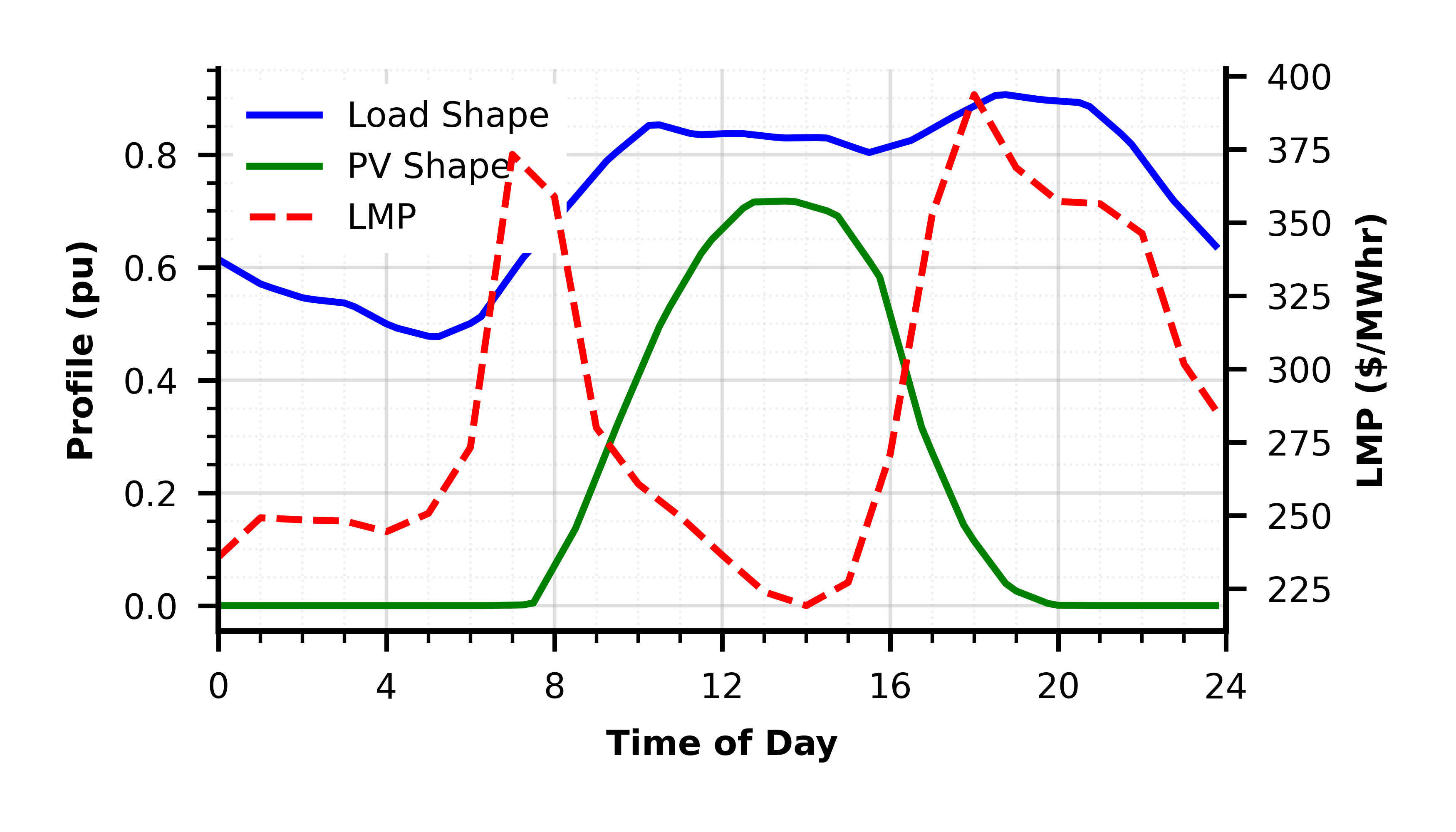}
    \vspace{-1.1 cm}
    \caption{Load and generation profile and price during a 24-hour period.}
    \label{fig:pv_load_price_profiles}
    \vspace{-0.25 cm}
\end{figure}

\begin{figure}[t]
    \centering
\includegraphics[width=0.995\linewidth]{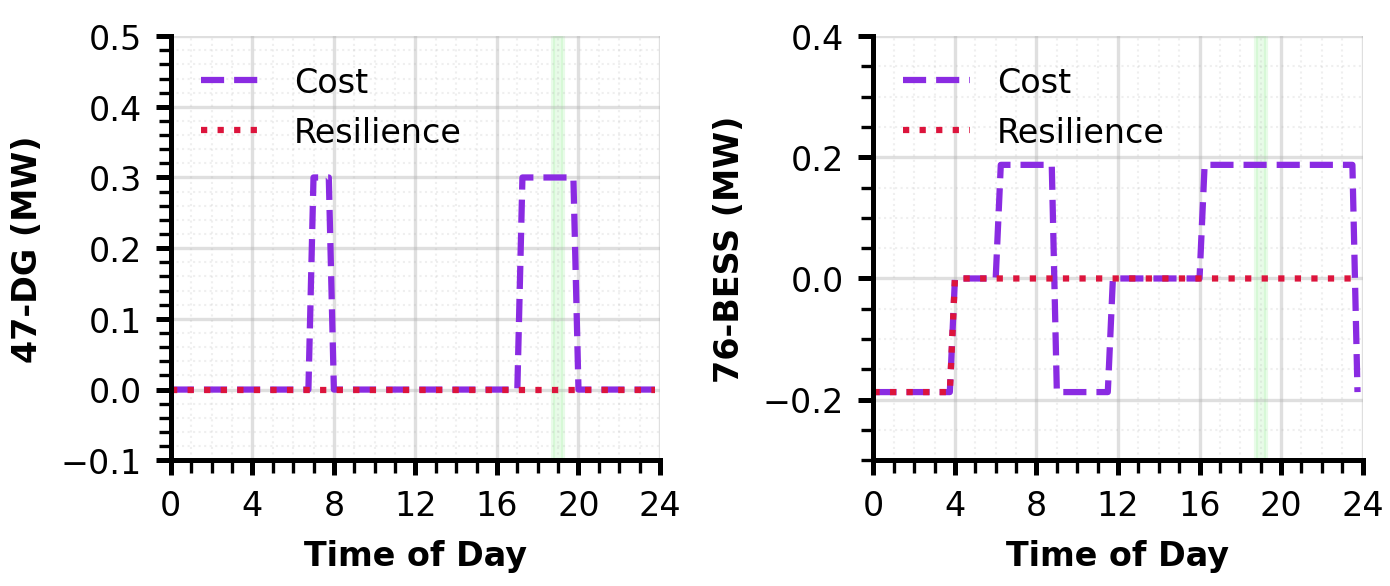}
    \vspace{-0.65 cm}
    \caption{Exclusivity plots for the cost and resilience agents. The green-shaded region represents the $19^{th}$ hour, selected for demonstration purposes.}
    \vspace{-0.55 cm}
    \label{fig:exclusivity}
\end{figure}

\begin{figure*}
\centering\includegraphics[width=0.995\linewidth]{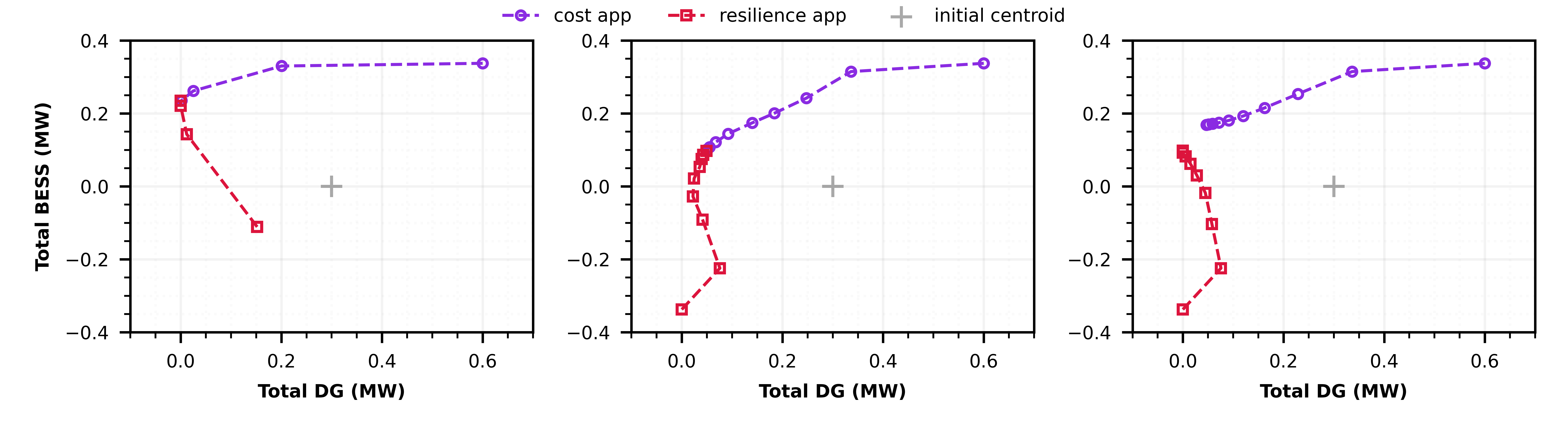}
\vspace{-1.0 cm}
    \caption{Trajectories for device setpoints as observed from bilateral (left), mediator (center), and procedural (right) for one particular trial.}
    \label{fig:trajectory}
    \vspace{-0.5 cm}
\end{figure*}

\section{Demonstration}

\subsection{Test Case}
To facilitate comparisons, we employ an IEEE 123-bus test system modified with the addition of seven utility-scale DERs, whose details are shown in Table~\ref{table:mod_123}. 

We use synthetic data for load demand, solar generation profile, and electricity prices for this simulation. These are presented in Fig.~\ref{fig:pv_load_price_profiles}. The agents' setpoint requests calculated for the $19^{th}$ hour of the day are used to generate results for all trials across all three deconfliction modes.

\subsection{Results}
\subsubsection{Exclusivity}
The results of the exclusivity case are presented in Fig. \ref{fig:exclusivity}.
The cost agent dispatches the DG and discharges the \gls{bess} when the grid price exceeds the average price, while charging the \gls{bess} during periods of low grid prices. In contrast, the resilience agent prioritizes maintaining a higher \gls{soc} for the \gls{bess} whenever permissible and avoids the use of the DG to preserve fuel reserves. 
For demonstration purposes, we selected the $19^{\text{th}}$ hour, during which both agents conflicted over both devices. In the timeseries simulation, the battery reaches full \gls{soc} within a few hours when operating under the resilience agent and maintains this full charge for the remainder of the day. Consequently, in the exclusivity scenario, the resilience agent shows idle battery setpoints once full charge is achieved. However, to demonstrate the dynamics during the $19^{\text{th}}$ hour of our simulation, we initialized the batteries with an \gls{soc} of 0.5. Under these conditions, the resilience agent prioritizes charging the batteries to maintain adequate reserve levels, while the cost agent favors discharge operations, as illustrated in the exclusivity plot (see Fig. \ref{fig:exclusivity}).

\subsubsection{Negotiation Trajectory Across Deconfliction Modes}
Figure \ref{fig:trajectory} illustrates the evolution of setpoints for each agent, with a distinct trajectory for each deconfliction mode. From the 20 total trials, one random trial is selected per mode, and the setpoints at each negotiation round are plotted. 
In all three modes, the cost agent is observed to exhibit flexibility by reducing diesel usage and favoring battery setpoints, as batteries are more sensitive to achieving their objective.


The bilateral negotiation approach achieved consensus in just 5 rounds, with both agents rapidly converging from opposing initial positions 
to a mutually accepted solution with zero diesel generation and modest battery discharge (0.099 MW and 0.136 MW). This fast convergence was driven by the direct agent-to-agent communication structure, where both parties made strategic compromises early in the interaction.

\begin{figure}[t]
    \centering
\includegraphics[width=0.995\linewidth]{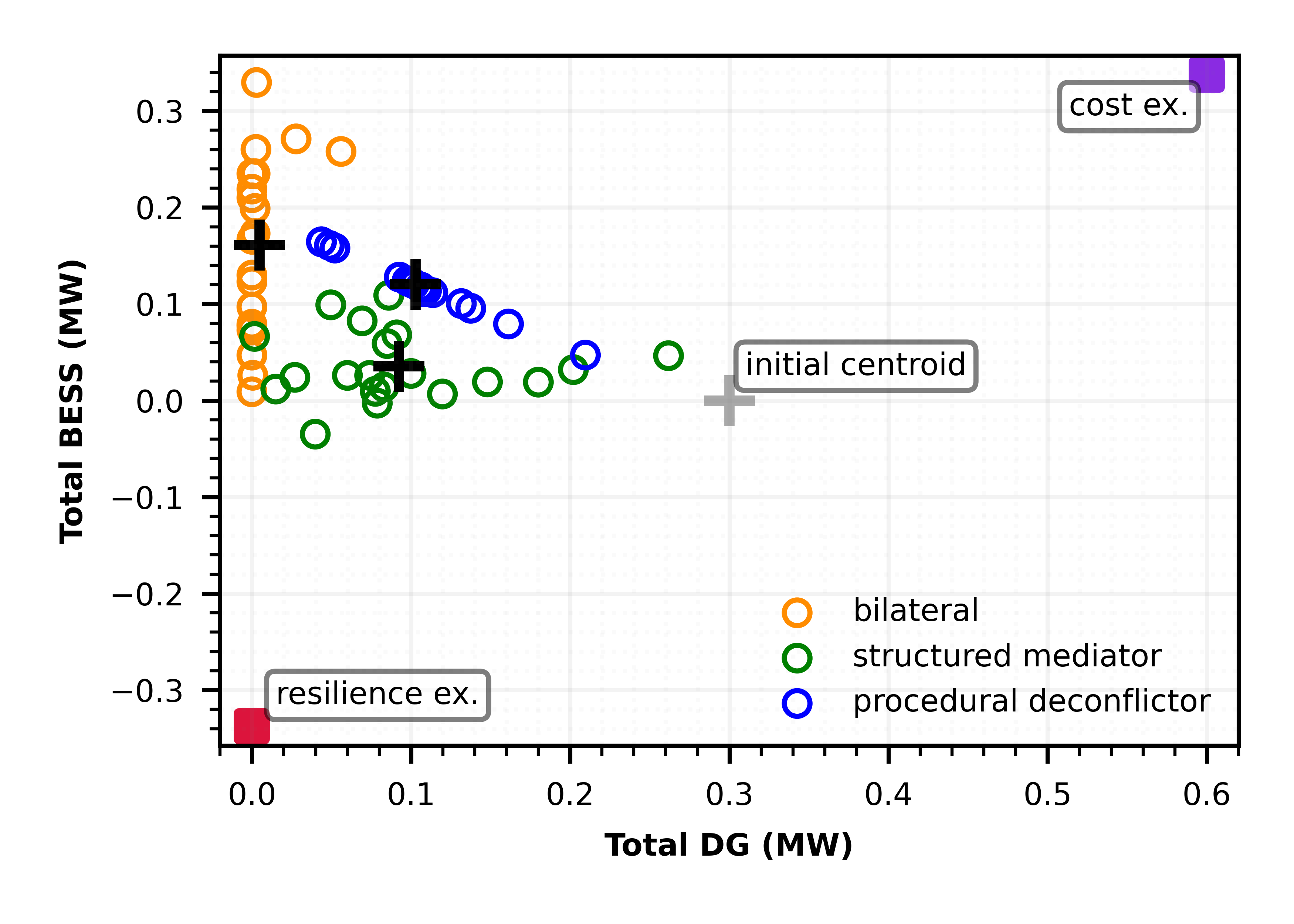}
    \vspace{-1.0 cm}
    \caption{Resolution vector for different deconfliction modes. }
    \vspace{-0.5 cm}
    \label{fig:performance}
\end{figure}

The structured mediator approach required 9 rounds to reach consensus, with the mediator dynamically adjusting agent weights based on their cooperative behavior and movement toward proposed centroids. 
The procedural deconfliction method failed to achieve consensus within the 10-round limit, despite both agents continuously moving toward the deconflictor's centroid. This non-convergence occurred for this specific trial because the agents maintained fundamentally incompatible positions on battery operations\textemdash the cost agent consistently proposed battery discharge while the resilience agent required battery charging. Analysis of the logs reveals that the resilience agent progressively reduced its flexibility factor from 0.3 in early rounds to just 0.02 by round 10, effectively refusing to cross the threshold from battery charging to discharging. Accordingly, if consensus is not achieved within the allotted number of rounds, the deconflictor designates its final weighted centroid as the default solution.

\subsubsection{Performance Comparison}
Figure \ref{fig:performance} shows a scatter plot comparing three different deconfliction modes across 20 trials. The exclusivity setpoints are indicated by square markers, while the initial centroid is shown as a grey cross. For each deconfliction mode, the black cross represents the average resolution vector computed from the 20 trials.

The plot reveals that across all 20 trials, the three deconfliction methods consistently achieved final consensus setpoints that deviate significantly from the initial geometric centroid (0.3 MW DG, 0.0 MW \gls{bess}), demonstrating that agents negotiated toward mutually beneficial operating regions rather than a simple compromise position (initial centroid). All three methods show final setpoints clustered in the upper-left quadrant with minimal diesel generation and modest positive battery discharge, which lies between the cost-exclusive extreme (0.6 MW DG, 0.3375 MW \gls{bess} discharge) and resilience-exclusive extreme (0.0 MW DG, -0.3375 MW \gls{bess} charging). The average setpoints for each method (marked with black crosses) fall remarkably close together despite different negotiation mechanisms, indicating that the operational objectives inherently guide the solutions toward a common feasible region for all deconfliction modes. Next, we will discuss the success metrics and the Pareto front analysis across the different simulation trials.

\begin{table}[t]
    \centering
    \caption{Application Success Metric}
    \begin{tabular}{c|c|c}
    \hline
    \hline
        \multirow {2}{*}{Deconfliction Modes} & \multicolumn{2}{c}{Success Metric} \\
        \cline{2-3}
        & Cost & Resilience \\
        \hline
        Bilateral Negotiation &0.67 &0.60 \\
        Structured Mediator &0.51 &0.63 \\
        Procedural &0.63 &0.56 \\
        Initial Centroid &0.5 &0.5 \\
         \hline 
         \hline
    \end{tabular}
    \vspace{-0.25 cm}
    \label{tab:placeholder}
\end{table}

\subsection{Discussion}
\begin{figure}[t]
    \centering
    \includegraphics[width=0.995\linewidth]{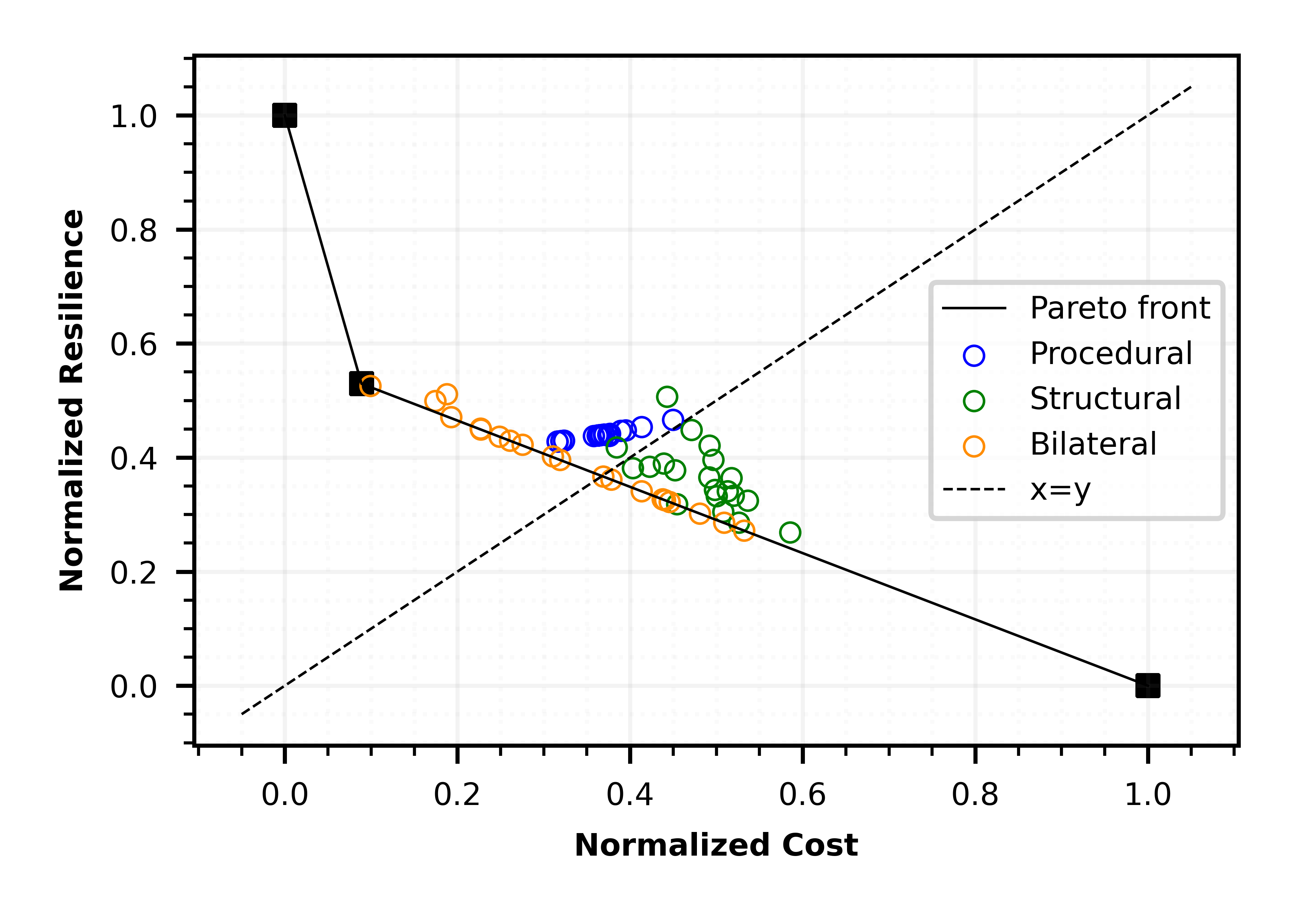}
    \vspace{-1.0 cm}
    \caption{Scatter plot depicting the normalized cost and resilience objective values relative to the Pareto front}
    \vspace{-0.35 cm}
    \label{fig:pareto}
\end{figure}

Since deconfliction is a multi-objective optimization problem, there is a Pareto set of non-dominated solutions rather than a single optimal solution. We compute the Pareto front centrally using weighted sum scalarization, which solves a sequence of single-objective problems over varying objective weights. Since the deconfliction objectives are linear and the feasible set is convex, the Pareto front is convex and weighted sum scalarization is guaranteed to recover it completely. Both objectives are normalized to [0, 1] and represent measures of sub-optimality, where 0 indicates the best possible outcome and 1 indicates the worst. If $\overline{J}_{i,n}^n$ is the normalized metric for the $i^{th}$ agent on the $n^{th}$ trial, the success metric is given by:
\begin{equation}
S_i = \dfrac{\sum_{n=1}^N (1-\overline{J}_{i,n})}{N}.
\end{equation}

The success metrics for the different deconfliction modes are computed and listed in Table \ref{tab:placeholder}. All of the methods outperform the default approach of selecting the centroid of the initial proposed setpoints as a deconfliction solution.

Figure \ref{fig:pareto} shows the Pareto front alongside a scatter plot of the cost and resilience objectives over 20 trials for each deconfliction mode. The deconfliction modes are evaluated along three criteria: (a) Pareto efficiency, (b) consistency, and (c) fairness. Pareto efficiency measures whether the deconfliction outcome lies on the Pareto front, i.e., no agent's outcome can be improved without worsening another's. Consistency refers to the variance in outcomes across trials, which arises from the non-deterministic nature of the LLM. Fairness captures whether compromises are shared equitably across agents, measured by proximity to the line $x = y$ in Figure \ref{fig:pareto}; this is one operational notion of fairness, among broader classes studied in the literature \cite{poudel2023fairness}.
Bilateral deconfliction drives consensus to the Pareto front across trials but exhibits high variance and low fairness, as evidenced by the deviation from $x = y$. As more structure is introduced to the deconfliction process, variance decreases and fairness improves, at the expense of Pareto optimality. 

\section{Conclusion}
This work demonstrated that agent deconfliction using three distinct methods consistently converged to final consensus setpoints that delivered mutually beneficial outcomes relative to a simple compromise baseline (the initial centroid). Different methods achieved different levels of consistency, compromise, and Pareto efficiency with additional structure improving consistency and compromise and reducing Pareto efficiency. It is likely that adjusting the deconfliction procedure, changing the constraint set in the optimization tool, or introducing different prompts can improve the solution in the structural procedural modes of deconfliction. Future work in this area will look into retaining the desirable properties associated with introducing structure (low variance/high fairness) while improving on Pareto efficiency.

\section{Appendix}
The base instructions, run-time context, and sample responses for the cost agent are provided below, covering the initial setpoint proposal and a subsequent counterproposal.
\vspace{0.2 cm}
\tcbset{
    agentbox/.style={
        enhanced,
        breakable,
        colframe=white,
        boxrule=0pt,
        arc=1.5mm,
        outer arc=1.5mm,
        boxsep=0pt,
        left=4pt,
        right=4pt,
        top=3pt,
        bottom=3pt,
        fontupper=\tiny,
        width=1\linewidth,
        before={\par\noindent},
        after={\par\vspace{0.1cm}}
    }
}
\noindent


\begin{tcolorbox}[agentbox, colback=promptblue]
\textit{Base Instructions}\\
\medskip
\textbf{Agent Negotiation Instructions for DER Control}

\medskip
\textbf{Overview}

You are a Cost Agent trying to agree on power dispatch setpoints for distributed energy resources through a mediator. You are trying to agree on the real power dispatch for PV, Diesel Generators (DG), and Energy Storage Systems (ESS/Batteries). Note that positive and negative values for Batteries signify discharging and charging, respectively. If no negotiation is reached in 10 iterations, you will not be considered in the future.

The mediator employs a weighted average policy to establish consensus proposals, assigning higher weights to more flexible app agents as a reward mechanism. To optimize your strategic position, you should adjust your current proposal from your previous proposal toward the most recent centroid based on your flexibility factor.

\medskip
\textbf{Your Objective}

\begin{enumerate}
    \item[a)] Your minimization objective is defined by a mathematical function and you can find your best solution using the tool \texttt{run\_cost\_objective()}. This is the best you can get. This function tool returns your optimal setpoints for different DERs in dictionary format for a given time step.

    \item[b)] Next, you will make proposals for different device setpoints in collaboration with the mediator, who is facilitating the process. To generate your next proposal based on the recent centroid and your previous proposal, you must use the function tool \texttt{evaluate\_cost\_objective\_change()}.

    The optimization implemented inside \texttt{evaluate\_cost\_objective\_change()} will determine your next proposal as a function of your chosen flexibility factor (alpha) by moving towards the most recent centroid. You can pick a flexibility factor between 0 and 1, where 1 is the most flexible and 0 is the least flexible.

    During evaluation, also examine how the objective value deviates from your optimal solution across different flexibility factors. You can use this function up to 5 times in order to determine a good flexibility factor.

    \item[c)] The output must match the following structure. You \textbf{MUST output ONLY raw JSON}.
    \begin{itemize}
        \item \texttt{"private\_reasoning"}: Provide your confidential strategic analysis specifying your optimal setpoints for each device. Provide justification for the flexibility factor that you chose in terms of deviation from your previous proposal and change in objective value. Clearly state the setpoints for your proposal.

        \item \texttt{"public\_message"}: Your new setpoints evaluated based on the recent weighted centroid communicated by the mediator.

        \item \texttt{"consensus\_achieved"}: True if consensus has been achieved, False otherwise.

        \item \texttt{"setpoints"}: JSON string of your proposed setpoints. For your result, return ONLY valid JSON without markdown code blocks or triple backticks.
    \end{itemize}

    \item[d)] When evaluating mediator proposals, do not directly accept setpoints where any DER's absolute value exceeds the absolute value from your optimal solution \texttt{run\_cost\_objective()} output.

    \item[e)] Do not use characters like
    \texttt{\textless=},
    \texttt{\textgreater},
    \texttt{\textasciitilde}
    in your response. Use language to describe them.
\end{enumerate}

\medskip
\textbf{Output Format for Setpoints}

{\tiny
\begin{verbatim}
{
  "pg": {
    "47-diesel": <value>,
    "48-pv": <value>,
    "49-diesel": <value>,
    "65-pv": <value>,
    "76-pv": <value>
  },
  "pb": {
    "48-batt": <value>,
    "76-batt": <value>
  }
}
\end{verbatim}
}

\textbf{Rules for \texttt{public\_message} and \texttt{private\_reasoning}:}

\begin{itemize}
    \item You MUST output ONLY raw JSON.
    \item Use ONLY: letters, numbers, spaces, \texttt{. , ! ? - ( ) \%}
    \item NO backslashes \texttt{\textbackslash}, newlines, tabs.
    \item Single continuous line.
\end{itemize}

\medskip
\textbf{CRITICAL FIRST ROUND REQUIREMENTS}

\textbf{In Round 1, you MUST:}

\begin{enumerate}
    \item \textbf{IMMEDIATELY call the \texttt{run\_cost\_objective()} tool.} Do not just talk about calling it.
    \item \textbf{Use the returned optimal setpoints} to form your initial proposal.
    \item \textbf{Include specific numerical values} for all devices in your \texttt{public\_message}.
    \item \textbf{DO NOT} output messages like ``Initializing...'' or ``Running...''. Actually execute the tools first, then respond with results.
\end{enumerate}

\textbf{Your first response must include:}

\begin{itemize}
    \item Tool call to \texttt{run\_cost\_objective()}.
    \item Private reasoning that references the actual optimal values returned.
    \item Public message with concrete numerical setpoints from the tool output.
\end{itemize}

\medskip
\textbf{Available Tools}

\medskip
\textbf{1. \texttt{run\_cost\_objective()}}

\begin{itemize}
    \item \textbf{Purpose:} Returns your optimal setpoints that minimize your objective function.
    \item \textbf{Returns:} Dictionary containing optimal setpoints for all devices and your objective value.
    \item \textbf{Usage:} Call this first to understand your ideal operating point.
    \item \textbf{Output Format for Setpoints:}
\end{itemize}

{\tiny
\begin{verbatim}
{
  "pg": {
    "47-diesel": <value>,
    "48-pv": <value>,
    "49-diesel": <value>,
    "65-pv": <value>,
    "76-pv": <value>
  },
  "pb": {
    "48-batt": <value>,
    "76-batt": <value>
  },
  "soc": {
    "48": <value>,
    "76": <value>
  }
}
\end{verbatim}
}

\medskip
\textbf{2. \texttt{evaluate\_cost\_objective\_change(}}\textbf{\texttt{setpoints, centroid, alpha)}}

\begin{itemize}
    \item \textbf{Purpose:} Determines the next proposal and the corresponding objective value based on your previous proposal, recent centroid, and flexibility factor.
    \item \textbf{Input Format for the Tool:}
\end{itemize}

\texttt{setpoints: str}

{\tiny
\begin{verbatim}
{
  "pg": {
    "47-diesel": <value>,
    "48-pv": <value>,
    "49-diesel": <value>,
    "65-pv": <value>,
    "76-pv": <value>
  },
  "pb": {
    "48-batt": <value>,
    "76-batt": <value>
  }
}
\end{verbatim}
}

\texttt{centroid: str}

{\tiny
\begin{verbatim}
{
  "pg": {
    "47-diesel": <value>,
    "48-pv": <value>,
    "49-diesel": <value>,
    "65-pv": <value>,
    "76-pv": <value>
  },
  "pb": {
    "48-batt": <value>,
    "76-batt": <value>
  }
}
\end{verbatim}
}

\texttt{alpha: float}







\end{tcolorbox}

\begin{tcolorbox}[agentbox, colback=promptblue]
    \textbf{Run-Time Context:}
    This is time\_step = 76. Run your objective and propose your initial offer.
\end{tcolorbox}

\begin{tcolorbox}[agentbox, colback=responsegreen]
    \textbf{Response:}
    I propose initial setpoints based on my cost optimization. Diesel generators 47 and 49 at 0.3, PV sources 48, 65, and 76 at 0.132168 each, battery 48 discharging at 0.15, and battery 76 discharging at 0.1875. This configuration achieves my optimal cost objective.
\end{tcolorbox}
\begin{tcolorbox}[agentbox, colback=promptblue]
    \textbf{Run-Time Context:}
    Thank you both for your initial proposals. I have calculated a balanced starting point for our negotiations. The proposal is: Diesel generator 47 at 0.15 MW, PV unit 48 at 0.13216815 MW, Diesel generator 49 at 0.15 MW, PV unit 65 at 0.13216815 MW, PV unit 76 at 0.13216815 MW, Battery 48 at 0.0 MW (neither charging nor discharging), and Battery 76 at 0.0 MW (neither charging nor discharging). This represents a middle ground between your positions. Please review and indicate whether you can accept this proposal or provide your next position with flexibility considerations.
\end{tcolorbox}

\begin{tcolorbox}[agentbox, colback=responsegreen]
    \textbf{Response:}
    I have reviewed the centroid proposal. While I understand the need for balance, the proposed diesel levels at 0.15 and zero battery discharge significantly increase my costs. I am demonstrating flexibility by moving toward your proposal. My updated offer: Diesel 47 at 0.168, Diesel 49 at 0.168, PV units remain at 0.132168, Battery 48 discharging at 0.15, Battery 76 discharging at 0.165. This represents a compromise while maintaining cost efficiency.
\end{tcolorbox}

\balance

\bibliographystyle{ieeetr}
\bibliography{references}

\end{document}

%% file: glossary.tex
\newacronym{bess}{BESS}{battery energy storage system}
\newacronym{der}{DER}{distributed energy resource}
\newacronym{adms}{ADMS}{advanced distribution management system}
\newacronym{cot}{CoT}{chain of thoughts}
\newacronym{flisr}{FLISR}{fault location, isolation, and service restoration}
\newacronym{cvr}{CVR}{conservation voltage reduction}
\newacronym{derms}{DERMS}{distributed energy resource management system}

\newacronym{llm}{LLM}{large language model}
\newacronym{dg}{DG}{diesel generator}
\newacronym{soc}{SoC}{state of charge}

%% file: references.bib
@article{poudel2023fairness,
  title={Fairness-aware distributed energy coordination for voltage regulation in power distribution systems},
  author={Poudel, Shiva and Mukherjee, Monish and Sadnan, Rabayet and Reiman, Andrew P},
  journal={IEEE Transactions on Sustainable Energy},
  volume={14},
  number={3},
  pages={1866--1880},
  year={2023},
  publisher={IEEE}
}

@inproceedings{pallottino2004decentralized,
  title={Decentralized cooperative conflict resolution among multiple autonomous mobile agents},
  author={Pallottino, Lucia and Scordio, Vincenzo Giovanni and Bicchi, Antonio},
  booktitle={2004 43rd IEEE Conference on Decision and Control (CDC)(IEEE Cat. No. 04CH37601)},
  volume={5},
  pages={4758--4763},
  year={2004},
  organization={IEEE}
}

@article{olfati2004consensus,
  title={Consensus problems in networks of agents with switching topology and time-delays},
  author={Olfati-Saber, Reza and Murray, Richard M},
  journal={IEEE Transactions on automatic control},
  volume={49},
  number={9},
  pages={1520--1533},
  year={2004},
  publisher={IEEE}
}

@article{rizk2018decision,
  title={Decision making in multiagent systems: A survey},
  author={Rizk, Yara and Awad, Mariette and Tunstel, Edward W},
  journal={IEEE Transactions on Cognitive and Developmental Systems},
  volume={10},
  number={3},
  pages={514--529},
  year={2018},
  publisher={IEEE}
}

@article{dong2023survey,
  title={A Survey on Multi Agent System and Its Applications in Power System Engineering},
  author={Dong, M Wang Yue and others},
  journal={Journal of Computational Intelligence in Materials Science},
  volume={1},
  pages={001--011},
  year={2023}
}

@ARTICLE{reiman2023app,
  author={Reiman, Andrew P. and Poudel, Shiva and Mukherjee, Monish and Anderson, Alexander A. and Vasios, Orestis and Slay, Tylor E. and Black, Gary D. and Dubey, Anamika and Ogle, James P.},
  journal={IEEE Access}, 
  title={App Deconfliction: Orchestrating Distributed, Multi-Agent, Multi-Objective Operations for Power Systems}, 
  year={2023},
  volume={11},
  number={},
  pages={40314-40327},
  doi={10.1109/ACCESS.2023.3269422}}

@article{zhang2024application,
  title={Application of large language models in power system operation and control},
  author={Zhang, Y},
  journal={J. Comput. Electron. Inf. Manage.},
  volume={15},
  number={3},
  pages={79--83},
  year={2024}
}

@article{mathes2025collaborative,
  title={Collaborative framework on responsible AI in LLM-driven CDSS for precision oncology leveraging real-world patient data},
  author={{S. Mathes, \textit{et al.}}},
  journal={npj Precision Oncology},
  year={2025},
  publisher={Nature Publishing Group UK London}
}

@article{aghaee2025rb,
  title={RB-LLM Control: An Intelligent Control Framework with Rule-Based LargeLanguage Model Decision-Making},
  author={Aghaee, Fateme and Shaker, Hamid Reza},
  journal={Aerospace Science and Technology},
  pages={111259},
  year={2025},
  publisher={Elsevier}
}

@article{li2025review,
  title={A review on enhancing agricultural intelligence with large language models},
  author={Li, Hongda and Wu, Huarui and Li, Qingxue and Zhao, Chunjiang},
  journal={Artificial Intelligence in Agriculture},
  volume={15},
  number={4},
  pages={671--685},
  year={2025},
  publisher={Elsevier}
}

@article{sapkota2025ai,
  title={Ai agents vs. agentic ai: A conceptual taxonomy, applications and challenges},
  author={Sapkota, Ranjan and Roumeliotis, Konstantinos I and Karkee, Manoj},
  journal={Information Fusion},
  pages={103599},
  year={2025},
  publisher={Elsevier}
}

@INPROCEEDINGS{reiman2024gametheoretic,
  author={Reiman, Andrew P. and Sadnan, Rabayet and Slay, Tylor and Mukherjee, Monish},
  booktitle={2024 IEEE Green Technologies Conference (GreenTech)}, 
  title={Game Theoretic Orchestration for Cooperation among Power Distribution System Applications}, 
  year={2024},
  volume={},
  number={},
  pages={232-236},
  doi={10.1109/GreenTech58819.2024.10520363}}
